\documentstyle[12pt]{article}
\normalbaselines
\begin{document}

\baselineskip=24pt

\bibliographystyle{unsrt}
\vbox {\vspace{6mm}}

\begin{center} {\bf BEYOND THE STANDARD "MARGINALIZATIONS" 
OF WIGNER FUNCTION}
\end{center}

\bigskip

\begin{center} 

{\it Stefano Mancini }$^{\dag}$
{\it Vladimir I. Man'ko }$^{\ddag}$   
{\it and Paolo Tombesi }$^{\dag}$

\end{center}

\bigskip

\begin{center}

$^{\dag}$ Dipartimento di Matematica e Fisica, Universit\`a 
di Camerino, 
I-62032 Camerino, Italy\\
$^{\ddag}$ Lebedev Physical Institute, Leninsky Prospekt 53, 117924 
Moscow, Russia

\end{center}

\bigskip

\bigskip

\begin{abstract}

\baselineskip=24pt

We discuss the problem of finding "marginal" distributions within 
different tomographic approaches to quantum state measurement,
and we establish analytical connections among them.
\end{abstract}

PACS number(s): 03.65.-w, 42.50.-p

\bigskip
\bigskip
\bigskip
\bigskip

\section{Introduction}\label{s1} 

\noindent

Recently the problem of measuring quantum states in quantum optics    
and quantum 
mechanics found experimental 
realization~\cite{raymer,wineland,mukamel,mlynek}.
Different procedures for measuring quantum 
states were suggested,
i.e., to measure an observable and from the experimental result of
the observations to reconstruct the density matrix of the quantum 
system
in terms, for example, of its Wigner function~\cite{wigner32}, or
its quasidistribution function like the Husimi 
$Q$-function~\cite{husimi40}
and the Glauber--Sudarshan 
$P$-distribution~\cite{glauber63,sudarshan63}.

The special interest to the problem of measuring the quantum states 
is 
related to the experimental realization of specific nonclassical 
quantum 
states~\cite{knight,drobny}. Such nonclasssical states, as one-mode
even and odd coherent states~\cite{physica74} (or Schr\"odinger cat 
states) were created in high-quality cavity (see, for 
example,~\cite{haroche}\,). It was also shown~\cite{gerry} that 
three-mode
even and odd coherent states maight be created in such experiments.
For an ion in a Paul trap, the even and odd coherent states 
may be realized~\cite{wineland,vogel1,nieto}, as well as the new 
type of
nonclassical states like nonlinear coherent 
states~\cite{f-osc,vogelnl}.

Thus, the problem of measuring such states, i.e., reconstructing 
their
density matrix in any representation, is actual. For trapped ion
several methods were indeed suggested for such a
reconstruction~\cite{schleich}.
Instead to determine the state of a radiation field 
there exists a dominant method, called optical tomography 
method~\cite{raymer}, which is
based on a relation of the Wigner function to the distribution 
function
of a homodyne observable~\cite{vogel,ber}, which may be derived from
the general scheme of $s$-ordered quasidistributions~\cite{cahgla}.

Recently, the optical tomography method was extended into symplectic
tomography ~\cite{mancini1}, in which the basic observable is the
generic squeezed and rotated quadrature. It was used to find 
classical-like description of the quantum dynamic 
evolution~\cite{mancini2}.
If in optical tomography the Wigner function is reconstructed
from measurable marginal distribution of the homodyne observable 
through 
Radon transform, in the symplectic tomography the Fourier 
transform is applied to reconstruct the density matrix by measuring 
the marginal distribution of the squeezed and rotated quadrature.
In this context, the symplectic tomography is similar to the 
field-strength method of Ref.~\cite{wvogel}. The symplectic 
tomography 
was generalized for multimode case as well~\cite{ariano}.

Recently, it was also suggested the photon number tomography 
method~\cite{walvogel,wodkie,mancini3}, for which the measurable 
observable
is the discrete photon number of the measurable field superposed 
with the local
oscillator field by scanning its complex amplitude. The photon 
number 
distribution, depending on the complex amplitude of the 
local
oscillator field, may be converted into the density matrix of the 
measurable 
field state. The symplectic tomography of nonclassical states
(namely, squeezed states and Schr\"odinger cat states) of an ion in
a Paul trap was discussed in Ref.~\cite{olga}. The reconstruction
of the density matrix of the squeezed vacuum, by using reproducible
measurements~\cite{landau} of the distribution for discrete photon 
numbers, was recently reported in Ref.~\cite{mlynek}.

Thus, in the theoretical and experimental framework of measuring
quantum states there exist different methods yielding the density
matrix from the corresponding measurable probability distributions.
On the other hand, till now the relations among all these  
distributions
were not clarified in details, to our knowledge. The aim
of this work is to establish the invertible transformations of the
symplectic 
quadrature distribution to the photon number distribution as well 
as to study the connection of the symplectic tomography scheme to 
the optical
tomography. Some aspects of this problem were discussed
in Ref.~\cite{JMO}. The relation of the Radon transform to the 
symplectic 
tomography procedure was discussed in Ref.~\cite{wuensche}. 
By finding
the explicit relations among the different measurable probability
distributions one is aimed to make prediction of measurements for
different observables, from measuruments, for example, of the 
marginal 
distribution of the generic squeezed and rotated quadrature in the
symplectic tomography scheme.

The paper is organized as follows. In Section~2, we review the 
properties of the marginal distributions 
extending the concept of "marginalizations" from the
optical tomography to the
symplectic, and photon number tomography. 
In Section~3, the connections among
the various "marginal" distributions are given. 
The main results obtained
are summarized in the Conclusion.

\section{Marginal Distributions}

\noindent

By refering to the standard definitions given in the 
literature~\cite{wigner32,ber}, by ``marginalization''
one should intend a line integral 
in the phase space $\{q,p\}$ of the 
Wigner function $W(q,p)$, i.e.
\begin{equation}\label{margdef}
w(x,\theta)=\int\frac{dqdp}{(2\pi)^2}
W(q,p)\delta(x-\cos\theta\,q-\sin\theta\,p)\,,
\end{equation}
where $\theta$ is the angle orientation of the line.
On the other hand, the marginal distribution $w(x,\theta)$ can
be considered as well as the Fourier transform of the 
characteristic
function $\chi(k)$
\begin{equation}\label{wchi}
w(x,\theta)=\int\frac{dk}{2\pi}\chi(k)e^{-ikx}\,,
\end{equation}
with
\begin{equation}\label{chi}
\chi(k)=\langle e^{ik(\cos\theta q+\sin\theta p)}\rangle\,.
\end{equation}
The marginal distribution represents a true measurable 
probability, at 
least in the field of optics, by means of homodyne detection 
methods.
Furthermore, recently has been shown~\cite{vogel} that from a 
collection
of marginal distributions it is possible to recover the phase 
space 
pseudo-distributions (tomography) or, even better, the density 
operator~\cite{APL} representing the quantum state of the system 
under study.

Our purpouse is now to generalize the previous definition of the 
marginalization in order to include all measurable probabilities 
related to observables which allow a tomographic reconstruction 
of the 
quantum state~\cite{JMO}.

In fact, the tomographic principle may be generalized as 
follow: given a density operator $\rho$ and a group element 
$\cal G$ (or to better say an 
operator belonging to
representation of a group, which acts in the space of quantum 
states), one can create different types of tomography if, by knowing 
the matrix elements
\begin{equation}\label{ins1}
w\left (y, {\cal G}\right )=
\langle y|{\cal G}\rho{\cal G}^{-1}|y\rangle
\end{equation}
from 
measurements, is able to invert the formula expressing the 
density 
operator in terms of the above ``marginal'' distribution 
$w\left (y, {\cal G}\right ),$ where $y$ may 
denote either continuous or discrete eigenvalues. 
The positive distribution $w\left (y, {\cal G}\right )$
is normalized
\begin{equation}\label{ins2}
\int w\left (y, {\cal G}\right )\,dy=1\,.
\end{equation}
It is to remark that, to simplify the terminology, from now on 
we refer 
indistincly to any positive, measurable, and normalized 
probability 
distribution like of Eq. (\ref{ins1}) as marginal distribution.

To get the density operator from a marginal distribution an 
inversion 
procedure should be employed, and one can use the properties 
of summation or 
integration over 
group parameters $\cal G$. 
Thus, the marginal distribution of the observable $y$ depends on 
extra
parameters determining the operator $\cal G$. It is worth  
noting that
the operator $\cal G$ may belong not only to a group but also 
to other
algebraic construction, for example, to quantum group. 
The only problem is mathematical 
one, to make 
the inversion, and/or physical one, to realize the 
transformation $\cal G$ the
in laboratory. 

Along this line the ``symplectic tomography'' has been 
introduced in Ref.~\cite{ariano} by means of the transformation
\begin{equation}\label{quadef}
\hat x=\mu \hat q+\nu \hat p
\end{equation}
with the real parameters $\mu,\,\nu$ generalizing the 
homodyne rotation. 
In this case the marginal distribution becomes 
\begin{equation}\label{margx}
w\,(x,\,\mu,\,\nu)=\int\frac{dq\,dp}{(2\pi)^2}
W(q,p)\delta(x-\mu q-\nu p)\,,
\end{equation}
while the density operator can be written in the invariant form as
\begin{equation}\label{density}
\hat\rho=\int dx\,d\mu \,d\nu ~w(x,\mu,\nu)\hat K_{\mu,\nu}\,,
\end{equation}
with the kernel operator 
\begin{equation}\label{ker}
\hat K_{\mu,\nu}=
=\frac{1}{2\pi}z^2e^{-izx}e^{-z(\nu-i\mu)\hat a^\dagger /\sqrt 2}
e^{z(\nu+i\mu)\hat a/\sqrt 2}e^{-z^2(\mu^2+\nu^2)/4}\,.
\end{equation}
Here $z$ represents the variable conjugate to $x$ by the 
Fourier transform.
The fact that $\hat K_{\mu,\nu}$ depends on the $z$ variable 
as well 
(i.e., each Fourier component gives a selfconsistent kernel)
 shows 
the overcompleteness of information achievable by measuring the
observable of Eq.~(\ref{quadef}). 
Due to this, one could set $z=1$ in~(\ref{ker}). 

Beside that, a ``photon number tomography'' was developed in 
Ref.~\cite{mancini3} parallely to Refs.~\cite{walvogel,wodkie}.
It is based on the possibility to measure the number of 
photons in the 
state displaced over the phase space. Hence, the marginal 
distribution 
will be 
\begin{equation}\label{P} 
w\left (n,\,\alpha \right )={\rm Tr}\,\{\hat 
D(\alpha)\hat\rho\hat
D^{-1}(\alpha)|n\rangle\langle n|\}
=\langle n|\hat D(\alpha)\hat\rho\hat 
D^{-1}(\alpha)|n\rangle\,,
\end{equation} 
and it also corresponds to a "propensity" \cite{Wod} obtained via 
filtering the original state described by ${\hat\rho}$ with the 
filter in 
a Fock state. The propensities have been also studied in 
connection with 
a quantum state reconstruction \cite{Buzek}. In some sense the 
probability $w(n,\alpha)$ differs from the other distribution 
$w(x,\mu,\nu)$ since it is intrinsically a phase space 
distribution (in 
the case of $n=0$ it corresponds to the Husimi-Q function). 
However, for 
convenience, we still refer to it as a marginal distribution 
in the sense 
previously explained.

The invariant expression for the density operator in terms of the 
distribution (\ref{P}) is
\begin{equation}\label{rhosample}
\hat\rho=\sum_{n=0}^{\infty}\int\frac{d^2\alpha}{\pi} 
\,w\left (n,\,\alpha \right )\hat K_s(n,\alpha)\,,
\end{equation} 
with the kernel given by
\begin{equation}\label{K}
\hat K_s(n,\alpha)=\frac{2}{1-s}
\left(\frac{s+1}{s-1}\right)^n\hat T(-\alpha,-s)\,.
\end{equation} 
The marginal distribution is normalized as
\begin{equation}\label{ins3}
\sum _{n=0}^\infty w\left (n,\,\alpha \right )=1\,.
\end{equation}
The 
operator $\hat T$ represents the complex Fourier transform of the 
$s$-ordered  displacement operator 
$\hat D(\xi,s)=\hat D(\xi)e^{s|\xi|^2/2}$,
which can also be written as~\cite{cahgla}
\begin{equation}\label{T2}
\hat T(\alpha,s)=\frac{2}{1-s}\,\hat 
D(\alpha)\left(\frac{s+1}{s-1}\right)^{\hat a^{\dag}\hat a}
\hat D^{-1}(\alpha)\,.
\end{equation} 

\section{Connection among marginal distributions}

Here we consider the possibility to express the above marginal 
distributions as function one of other.

First, since the operator~(\ref{ker}) 
is proportional to the displacement operator generating the 
coherent state 
from the vacuum, one can use the known matrix elements of 
the displacement
operator in the Fock basis, expressed in terms of Laugerre 
polinomials~\cite{cahgla}, to get the density matrix in the 
photon number
basis as a convolution of the magrinal distribution 
$w\,(x,\,\mu,\,\nu)$
\begin{equation}\label{A}
\langle m|\rho |n\rangle =\sqrt {\frac {n!}{m!}}
\frac {2^{(n-m)/2}}{2\pi }\int e^{ix-(\mu ^2+\nu ^2)/4}
w(x,\mu ,\nu )(\nu -i\mu )^{m-n}L_n^{m-n}
\left (\frac {\nu ^2
+\mu ^2}{2}\right )dx\,d\mu \,d\nu \,;~~~m>n\,.
\end{equation}
For $m<n\,,$ we use the hermiticity property of the 
density matrix 
$\langle m|\rho |n\rangle =\langle n|\rho |m\rangle ^*.$
The diagonal matrix elements of the density matrix in the 
Fock basis
give the photon distribution in terms of the marginal 
distribution
\begin{equation}\label{AA}
\langle n|\rho |n\rangle =P(n)=
\frac {1}{2\pi }\int e^{ix-(\mu ^2+\nu ^2)/4}\,
w(x,\mu ,\nu )\,L_n\left (\frac {\nu ^2
+\mu ^2}{2}\right )dx\,d\mu \,d\nu \,.
\end{equation}
Let us now rewrite Eq.~(\ref{P}) as
\begin{equation}\label{wnrho}
w\left (n,\,\alpha \right )=\sum_{k,l}\langle n|D(\alpha)|k\rangle
\langle k|\rho|l\rangle\langle l|D^{-1}(\alpha)|n\rangle\,,
\end{equation}
then we can use the density matrix elements of Eq. 
(\ref{A}) and the 
number representation of the displacement operator 
given in Ref.~\cite{cahgla}, to get
\begin{eqnarray}\label{C}
w\left (n,\,\alpha \right )&=&\frac {1}{2\pi }\int 
w\,(x,\mu ,\nu )
\exp \left \{ix-\frac {\mu ^2 +\nu ^2}{4}
+\frac {\alpha (\nu +i\mu )}{\sqrt 2}
-\frac {\alpha ^* (\nu -i\mu )}{\sqrt 2}\right \}\nonumber\\
&&\times L_n\left (\frac {\mu ^2+\nu ^2}{2}
\right )\,dx\,d\mu \,d\nu \,.
\end{eqnarray}
For $\alpha =0\,,$ Eq.~(\ref{C}) gives 
$w\left (n,\,0\right )=P(n)\,,$ where
$P(n)$ is the photon distribution~(\ref{AA}).

 To derive the inverse 
relation, we start
with an oscillator in the thermal equilibrium state. 
One can easily obtain
the quadrature marginal distribution for the thermal 
equilibrium state of 
the harmonic oscillator, which is interacting with 
a heat bath at dimensionless
temperature $T,$ in the form
\begin{equation}\label{a}
w_T(x,\,\mu ,\,\nu )=\left [\pi\,(\mu ^2+\nu ^2)\,\coth \,
[(2T)^{-1}]\right ]^{-1/2}\,\exp 
\left [-\frac {x^2}{(\mu ^2+\nu ^2)
\,\coth \,[(2T)^{-1}]}\right ].
\end{equation} 
It means that for the non normalized state density 
operator of the
harmonic oscillator of the form
\begin{equation}\label{b}
\hat \rho _{\rm {non}}=\exp 
\left (-\frac {a^\dagger a}{T}\right )
\end{equation}
we have the corresponding non normalized marginal 
distribution of the form
$~(\hat \rho _{\rm {non}}\rightarrow w_{\rm {non}})~$
\begin{equation}\label{c}
w_T^{\rm {non}}(x,\,\mu ,\,\nu )=\frac {1}{1-\exp \,(-1/T)}\,
w_T(x,\,\mu ,\,\nu )\,.
\end{equation} 
If the density operator corresponds to the shifted 
equilibrium position
in the driven oscillator phase space, i.e.,
\begin{equation}\label{d}
\hat \rho _{\rm {shift}}=D(\alpha)\,
\hat \rho _{\rm{non}}D(-\alpha )=
\exp \left (-\frac {(a^\dagger-\alpha ^*)\,
(a-\alpha )}{T}\right ),
\end{equation}
we have the non normalized marginal distribution of 
the driven oscillator
\begin{equation}\label{e}
w_{\alpha,s}(x,\,\mu ,\,\nu )=w_T^{\rm {non}}\left (x-
\mu \,\frac {\alpha +\alpha ^*}{\sqrt 2}-
\nu \,\frac {\alpha -\alpha ^*}{i\,\sqrt 2},\,
\mu ,\,\nu \right )\,.
\end{equation} 
One can see that the temperature $T$ is
\begin{equation}\label{f}
T=\left [\ln\,\frac {s-1}{s+1}\right ]^{-1}
\end{equation} 
and the operator $[(1-s)/2]\,\hat T(\alpha ,\,s)$ 
has exactly the form 
of the operator~(\ref{d}). It means that the 
non normalized marginal 
distribution corresponding to the operator 
$\hat T(\alpha ,\,s)$ is
\begin{eqnarray}\label{g}
w_{\alpha ,\,s}(x,\,\mu ,\,\nu )
&=&\frac {1-s}{2}\,\frac {1}{1
-\exp \,(-1/T)}\left [\pi\,(\mu ^2+\nu ^2)\,\coth \,
[(2T)^{-1}]\right ]^{-1/2}\nonumber\\
&\times &\exp \left \{-\frac {\left (x-\left[\mu \,(\alpha 
+\alpha ^*)/\sqrt 2\right ]-\left[\nu \,(\alpha 
-\alpha ^*)/(i\,\sqrt 2)\right ]\right )^2}{(\mu ^2+\nu ^2)\,
\coth \,[(2T)^{-1}]}\right \},
\end{eqnarray} 
where the temperature $T$ is given by~(\ref{f}). 
Expression~(\ref{g})
gives the connection of the marginal distribution 
$w\left (x,\,\mu ,\,\nu \right )$
with the photon number marginal distribution 
$w\left (n,\,\alpha \right )$. 
In fact, by taking the expectation value of both side of Eq. 
(\ref{rhosample}) over the eigenket $|x\rangle$ of the quadrature 
operator (\ref{quadef}), one obtains
\begin{equation}\label{h}
w\,(x,\,\mu ,\,\nu )=\int \frac {d^2\alpha }{\pi}\,
\sum_{m=0}^\infty \frac {2}{1-s}\left (\frac {s+1}{s-1}\right )^m
\,w\left (m,\,\alpha \right )w_{-\alpha ,\,-s}(x,\,\mu ,\,\nu )\,,
\end{equation}
where
\begin{eqnarray}\label{m}
w_{-\alpha ,\,-s}(x,\,\mu ,\,\nu )&=&\frac {1+s}{2}\,\frac {1}{1
-\exp \,(-1/T)}\left [\pi \,(\mu ^2+\nu ^2)\,\coth \,
[(2T)^{-1}]\right ]^{-1/2}\nonumber\\
&\times &\exp \left \{-\frac {\left (x+\left[\mu \,(\alpha 
+\alpha ^*)/\sqrt 2\right ]+\left[\nu \,(\alpha 
-\alpha ^*)/(i\,\sqrt 2)\right ]\right )^2}{(\mu ^2+\nu ^2)\,
\coth 
\,[(2T)^{-1}]}\right \}
\end{eqnarray} 
and the ``temperature'' $T$ is given by~(\ref{f}) with 
$s\rightarrow -s\,,$ i.e.,
\begin{equation}\label{n}
T=\left (\ln \frac {s+1}{s-1}\right )^{-1}.
\end{equation}
The equation~(\ref{h}), along with~(\ref{m}) and (\ref{n}), is the 
inverse 
relation for~(\ref{C}). Thus, measuring the quadrature marginal 
distribution
$w\,(x,\,\mu,\,\nu)$ in homodyne experiments one can predict 
the results 
of measurement of the photon marginal distribution and, 
vice versa,
measuring the photon marginal distribution one can predict 
the results
of measurements of the quadrature marginal distribution.

Let us now address the question how to find the rotated 
quadrature 
marginal distribution $w\left (x,\,\theta \right )$
of the optical tomography scheme if one knows the squeezed and 
rotated quadrature distribution $w\left (x,\,\mu ,\,\nu  \right )$
of symplectic tomography. The answer is obvious, namely,
\begin{equation}\label{lis1}
w\left (x,\,\cos \,\theta ,\,\sin \,\theta \right )
=w\left (x,\,\theta \right )\,.
\end{equation}
Formula~(\ref{lis1}) may be rewritten in the integral form
\begin{equation}\label{lis4}
w\left (x,\,\theta \right )=\int w\left (x,\,\mu ,\,\nu \right )
\delta \left (\mu -\cos \,\theta \right )\,
\delta \left (\nu -\sin \,\theta \right )\,d\mu \,d\nu\,.
\end{equation}
So, given the function $w\left (x,\,\mu ,\,\nu  \right )$
one gets the homodyne marginal distribution
$w\left (x,\,\theta \right ).$
The inverse problem may also be solved, namely, how to find 
the marginal distribution of the symplectic tomography 
$w\left (x,\,\mu ,\,\nu  \right ),$ if one knows the homodyne
marginal distribution $w\left (x,\,\theta \right ).$
It means that knowing the distribution function of two
variables one has to reconstruct the distribution function of 
three variables. 

To make the reconstruction, we first decompose the periodic 
function $w\left (x,\,\theta \right )$ into the Fourier
series
\begin{equation}\label {lis2}
w\left (x,\,\theta \right )=\sum _{n=0}^\infty
\left[c_n\left (x\right )\cos \,n\theta 
+d_n\left (x\right )\sin \,n\theta\right]\,.
\end{equation}
Then, in view of the known espressions for $\cos \,n\theta \,,
\sin \,n\theta $ as polynomials in $\sin \,\theta \,,
\cos \,\theta \,,$ we make in~(\ref{lis2}) the replacement 
$$\cos \,\theta \rightarrow \mu \,;\sin \,\theta \rightarrow 
\nu \,.$$
Thus, we get the expression of the marginal distribution 
$w\left (x,\,\mu ,\,\nu  \right )$ in the form
\begin{equation}\label{lis3}
w\left (x,\,\mu ,\,\nu \right )=\frac {1}{2\pi}\int _0^{2\pi}
w\left (x,\,\theta \right )\,d\theta +\frac {1}{2\pi}
\sum _{n=1}^\infty \left [\left (\mu +i\nu \right )^n
\int _0^{2\pi }e^{-in\theta }\,w\left (x,\,\theta \right )
\,d\theta +c.c.\right ].
\end{equation}
Finally, by knowing the marginal distribution of the optical 
tomography 
scheme we directly obtain the marginal distribution of symplectic
tomography.

Using~(\ref{h}) one can find the connection of the photon number 
tomography 
to the optical tomography formalism. The marginal 
distribution of the 
rotated quadrature is expressed in terms of the photon 
number marginal
distribution as follows
\begin{equation}\label{lis5}
w\left (x,\,\theta \right )=\int \frac {d^2\alpha }{\pi}\,
\sum_{m=0}^\infty \frac {2}{1-s}\left (\frac {s+1}{s-1}\right )^m
\,w\left (m,\,\alpha \right )w_{-\alpha ,\,-s}\left (x,\,
\cos \,\theta ,\,\sin \,\theta \right )\,,
\end{equation}
where
\begin{eqnarray}\label{lis6}
w_{-\alpha ,\,-s}\left (x,\,
\cos \,\theta ,\,\sin \,\theta \right )
&=&\frac {1+s}{2}\,\frac {1}{1
-\exp \,(-1/T)}\left [\pi \coth \,
[(2T)^{-1}]\right ]^{-1/2}\nonumber\\
&\times& \exp \left \{-\frac {\left (x+\left[\cos \,\theta 
\,(\alpha +\alpha ^*)/\sqrt 2\right ]+\left[\sin \,\theta 
\,(\alpha -\alpha ^*)/(i\,\sqrt 2)\right ]\right )^2}{\coth 
\,[(2T)^{-1}]}\right \}\nonumber\\
\end{eqnarray} 
and the ``temperature'' $T$ is given by~(\ref{n}).

\section{Conclusion}

\noindent

We have provided analytical relations which connect the "marginal" 
distributions of different tomographic processes. Physically 
they will 
give predictions on how to make cross checking of the state 
measurements or how, by
using one sort of measurement, to reconstruct another sort, 
i.e., by 
measuring an observable one gets information on the 
distribution of other 
observables.

\end{document}